\documentclass[a4paper,twoside]{article}

\usepackage{epsfig}
\usepackage{subcaption}
\usepackage{calc}
\usepackage{amssymb}
\usepackage{amstext}
\usepackage{amsmath}
\usepackage{amsthm}
\usepackage{multicol}
\usepackage{pslatex}
\usepackage{apalike}
\usepackage{balance}
\usepackage{hyperref}
\usepackage{SCITEPRESS}     %

\newcommand{\tthref}[2]{\href{#1}{\texttt{#2}}}

\usepackage{tikz}
\usepackage[some,bottom]{background} 
\newcommand\copyrighttext{
 \footnotesize
  This is the author's version of the work. It is posted here for your personal use.
  
  \vspace{0.5em}
  
  The definitive Version of Record was published in 2019 by SciTePress in \emph{Proceedings of the 14th International Conference on Software Technologies (ICSOFT 2019)}, pages 552-559, ISBN: 978-989-758-379-7, doi: \tthref{https://doi.org/10.5220/0007967905520559}{10.5220/0007967905520559} (CC BY-NC-ND 4.0).
 }
\newcommand\copyrightnotice{%
\begin{tikzpicture}[remember picture, overlay]
\node[yshift=10pt]
{\fbox{\parbox{\dimexpr\textwidth-\fboxsep-\fboxrule\relax}{\copyrighttext}}};
\end{tikzpicture}
}
\backgroundsetup{opacity=1, scale=1, angle=0, contents={
\copyrightnotice
}}
\SetBgOpacity{1}
\SetBgScale{1}
\SetBgAngle{0}
\SetBgContents{\copyrightnotice}
\SetBgColor{black}

\begin{document}

\title{Towards Using Data to Inform Decisions in Agile Software Development: Views of Available Data}

\author{\authorname{Christoph Matthies\sup{1}\orcidAuthor{0000-0002-6612-5055}, Guenter Hesse\sup{1}\orcidAuthor{0000-0002-7634-3021}}
\affiliation{\sup{1}Hasso Plattner Institute,\\University of Potsdam, Germany}
\email{christoph.matthies@hpi.de, guenter.hesse@hpi.de}
}


\abstract{
Software development comprises complex tasks which are performed by humans.
It involves problem solving, domain understanding and communication skills as well as knowledge of a broad variety of technologies, architectures, and solution approaches.
As such, software development projects include many situations where crucial decisions must be made.
Making the appropriate organizational or technical choices for a given software team building a product can make the difference between project success or failure.
Software development methods have introduced frameworks and sets of best practices for certain contexts, providing practitioners with established guidelines for these important choices.
Current Agile methods employed in modern software development have highlighted the importance of the human factors in software development.
These methods rely on short feedback loops and the self-organization of teams to enable collaborative decision making.
While Agile methods stress the importance of empirical process control, i.e. relying on data to make decisions, they do not prescribe in detail how this goal should be achieved.
In this paper, we describe the types and abstraction levels of data and decisions within modern software development teams and identify the benefits that usage of this data enables.
We argue that the principles of data-driven decision making are highly applicable, yet underused, in modern Agile software development.

}

\onecolumn \maketitle \normalsize \setcounter{footnote}{0} \vfill
\BgThispage

\section{\uppercase{Introduction}}
The practice of software development includes series of decisions that must be made to ensure the success of a project.
These decisions concern not only the scope, budget and feature set of the product being developed, but also how development teams are organized, what technologies and architectures are employed, how the customer is engaged and how requirements are elicited and prioritized.
Making the right decisions for a given context, i.e. the decisions that have the highest chance of leading to project success, is of critical importance~\cite{Drury2012}.
Failing to do so is likely to result in overruns of budget and schedule, lost opportunities for the organization in need of the developed software, and ultimately project cancellation~\cite{Taherdoost2018,Molokken2005}.
As software is becoming pervasive and is being employed in all facets of life and in a large variety of contexts, few universal rules for decisions in software development, applicable to most situations and conditions, can be defined~\cite{kuhrmann2018}.

Modern software projects must be flexible, adapting to changing requirements and circumstances by analyzing available data, and making appropriate decisions.
In Agile methods, such as Scrum and Kanban~\cite{Reddy2015}, this idea is realized through short, iterative feedback cycles~\cite{Williams2003}, relying on the capabilities of team members in self-organizing teams~\cite{Hoda2010,Matthies2016c} and making use of available data to enable ``evidence-based decision making''~\cite{Fitzgerald2014}.
The Scrum Guide states: ``Scrum is founded on empirical process control theory [...] knowledge comes from experience and making decisions based on what is known. [...]''~\cite{Schwaber2017}.
However, few details are given as to the concrete implementations of these concepts or what data and knowledge is available and relevant to teams employing Agile software development approaches.
In this paper, we provide an overview of the described approaches that make use of data in informing decisions in Agile software development and highlight challenges and possible conflicting aspects.
Data-informed decision-making approaches are valuable for organizing and administering software projects.
Through a better understanding of their role and impact, they can be applied more thoroughly in the future.

\section{\uppercase{Data-Driven Decision Making}}
Different approaches have previously been proposed for using data to inform and support individuals and project teams in making the most likely correct decisions.
Related work reaching back multiple decades includes frameworks and theories that investigate this area.
The terms \emph{Data-Driven Decision Management}, \emph{Data-Directed Decision Making} and \emph{Data-Driven Decision Making} (DDDM) have been used to refer to the practice of basing decisions on the analysis of collected data.

\subsection{Application of DDDM}
The application of DDDM concepts is contrasted with the process of coming to a decision by relying on ``gut feeling'', personal experience and intuition~\cite{Brynjolfsson2011,Provost2013}, which, in the absence of capable analysis technologies, was the de facto standard throughout the history of commercial enterprises.
One of the core ideas of DDDM is that decisions, as well as their estimated efficacy, can be deduced from key data sets.
In a basic example, an employee in marketing, tasked with creating and selecting advertisements to be shown to website visitors, could base their designs and selections solely on their long experience of working in the advertising field and their intuitive understanding of the effects of different marketing campaigns.
However, when applying DDDM, the employee could additionally analyze how website users have interacted with ads in the past in order to draw conclusions for the future.
In line with our arguments, Provost and Fawcett point out that these two approaches are not mutually exclusive and can be combined~\cite{Provost2013}.

Technology adoption in business scenarios has considerably increased in recent years, due to the continuing digital transformation~\cite{Hesse2018} and the reliance on Internet technologies~\cite{Afuah2002}.
DDDM has, therefore, become an influential part of a large variety of industries~\cite{Provost2013}, including highly important sectors such as medicine~\cite{hollis2015}, transportation, and manufacturing~\cite{Hesse2019}.
Studies of the application of DDDM in companies have shown the benefits of the approach.
Brynjolfsson et al. collected data on the business practices and information technology investments of 179 businesses and correlated this data with company performance measures such as productivity, profitability, and market value~\cite{Brynjolfsson2011}.
The authors find that companies that adopted DDDM show an increase of 5-6\% regarding output and productivity than what would be expected given their investments and information technology usage.

While the introduction of digital networks and communication infrastructure in enterprises has improved efficiency, it has also intensified existing tensions and led to new challenges.
Technology has enabled greater transparency and visibility throughout businesses.
Stakeholders, e.g. customers or internal users, have instantaneous access to actionable information on the state of projects and real-time situational awareness is dramatically increased~\cite{Schrage2016}.
Therefore, the managerial and operational ability to act on the collected data and associated analyses must also scale.

\subsection{Being Driven or Being Informed by Data}
Recent work in the field of decision support for software projects has focused on \emph{data-driven} approaches~\cite{Svensson2019,Olsson2014,Provost2013}.
This term highlights the fact that data is in the idiomatic ``driver's seat'', being responsible for the direction a project is headed in.
This approach is in direct opposition to relying solely on intuition and the experience of team members, which are classified as biased and unreliable.
However, software engineering is still a quintessentially human task~\cite{FernandoCapretz2014}, requiring creativity, problem-solving abilities, and empathy for the users and stakeholders who will eventually use, or be impacted by, the developed software.
Disregarding or assigning little value to this aspect, i.e. concentrating on only the data that is produced by teams and software users, ignores the value that attention to human factors can provide in a project~\cite{Biddle2018a,Sherdil}.
We, therefore, propose being \emph{informed} instead of \emph{driven} by data concerning decision making in the context of software development, taking into account analyses of the available data and highlighting the need for human interpretation.
While important decisions that can affect project success should not only be based on intuition, they should also not solely rely on data, which may likewise only portray a one-sided, biased view.
Instead, as much project data as possible should be collected and analyzed, which can then be interpreted \emph{or ignored as irrelevant} by practitioners and team members that have intricate knowledge of the data and the context in which it was gathered.
This applies not only to software developers, who are experts concerning the developed application and the technical details, but also product managers, sales departments, and other roles that have knowledge related to the user and the context.
Combining data and human interpretation can enable better \emph{informed} decisions in software development.

\section{\uppercase{Agile Software Methods}}
Agile development methods are closely tied to the availability of data.
It enables teams to implement the short feedback cycles that define ``agility''~\cite{Schwaber2017}.

\subsection{Evolution of Agile Software Development}
Iterative and incremental development (IID) approaches for building software are not new.
Larman and Basili trace these concepts back to the 1930s, identifying the 1970s and 80s as the most active but least known part of their history~\cite{Larman2003}.
They describe the evolution of Agile methods, starting with the Rational Unified Process~\cite{kruchten2004} and the Dynamic Systems Development Method~\cite{stapleton1997dsdm} from the 1990s.
Several key ideas followed, such as Extreme Programming~\cite{Beck2000}, the Crystal family of methods~\cite{cockburn2004crystal} and finally Feature-Driven Development~\cite{palmer2001practical}, Lean~\cite{poppendieck2003lean} and Scrum~\cite{Schwaber2017}

Agile software development methods represent a set of best practices for software development in teams, created by experienced practitioners~\cite{fowler2001agile,Dyba2008}.
These methods highlight communication, adaptation to change, innovation, and teamwork, emphasizing productivity rather than process rigor~\cite{Agerfalk2006}.
They are often portrayed as the antithesis to traditional, more plan-based approaches, which feature rationalized and planned decision making, with work progressing in planned, successive stages, with little feedback built into the system~\cite{Lei2017}.
Williams and Cockburn claim that software development cannot be considered a ``defined process'', as change is inevitable, and requirements will be adjusted during the time the product is being developed~\cite{Williams2003}.
Software development is instead recognized as a flexible and ``empirical (or nonlinear)'' process~\cite{Williams2003}.
In the context of engineering software, development processes, therefore, require short ``inspect-and-adapt'' cycles and continuous feedback loops to enable process improvements based on empirical evidence.
These ideas are part of the \emph{Agile Manifesto}~\cite{fowler2001agile}, written by the practitioners who proposed many of the modern Agile development methods.
It presents four core values, among them ``responding to change over following a plan''.
The importance of adapting to changing circumstances through feedback loops is also highlighted in the accompanying Agile principles, which state that ``at regular intervals, the team reflects on how to become more effective, then tunes and adjusts its behavior accordingly''~\cite{fowler2001agile}.
The Scrum Guide~\cite{Schwaber2017}, the seminal text for the currently most popular Agile development method~\cite{VersionOne2018}, likewise incorporates these concepts, stating that ``transparency, inspection, and adaptation'' are prerequisites for making decisions in teams based on data.

\subsection{Agile Decision-Making}
\begin{figure}[htbp]
    \includegraphics[width=\columnwidth]{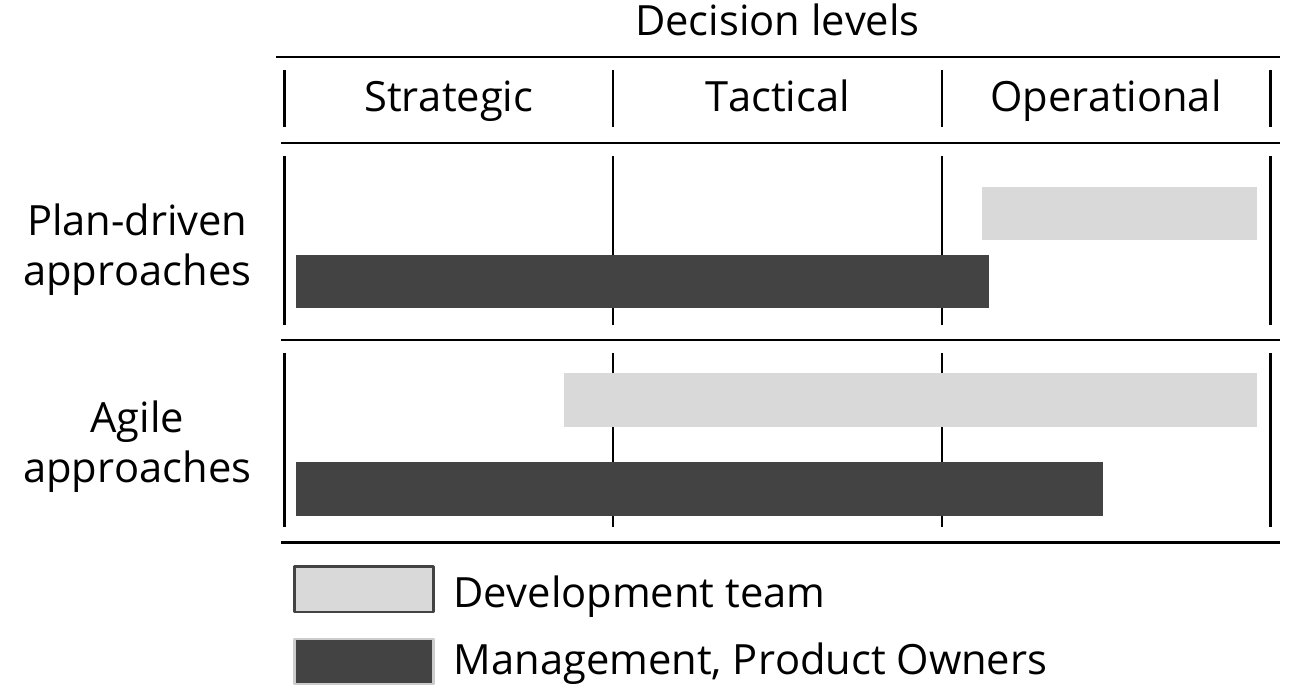}
    \caption{Differences between plan-driven and Agile approaches regarding responsibilities of different levels of decisions, adapted from Moe et al.~\cite{Moe2012}.}
    \label{fig:decision_levels}
\end{figure}

The shift from a more plan-driven approach to Agile development methods in an organization also requires a change in how decisions are made~\cite{Moe2012}.
It implies converting classical ``command and control'' attitudes to approaches relying on shared decision-making involving stakeholders and development teams~\cite{moe2009overcoming}.
A central point of Agile methods is the focus on working in small teams that have access to customers as well as the product's users and those affected.
The responsibility of establishing the priority of features and work items falls jointly on the development team and the stakeholders, who have different backgrounds and goals~\cite{nerur2005challenges}.
This shared, flexible, decision-making represents a ``barely sufficient'' command and control approach~\cite{mcavoy2009role}.
The decision-making process in Agile methods has been described as \emph{naturalistic}, compared to the \emph{rational} decision-making of plan-driven approaches~\cite{Moe2012}.
In these frameworks, a rational decision complies with a set of rules that govern behavior, which are applied in a logical fashion to generate decisions with acceptable consequences for the decision makers.
Rational decisions are based on the assumption that the set of solution possibilities and the probability of outcomes is known~\cite{Zannier2007}.
In comparison, a naturalistic decision-making process is characterized by situational behavior, the diminished importance of conscious analytical evaluation and application of context-dependent rules~\cite{klein2008naturalistic}.
The principles of Agile software development more closely align with the definition of naturalistic decision-making, as the importance of awareness and project context are highlighted~\cite{Zannier2007}.
The types of decisions and plans that have to be made in a software business can be split into three main domains: strategic, tactical, and operational.
Figure~\ref{fig:decision_levels} gives an overview of these levels and their assignments to different team roles in plan-driven and Agile software development approaches.

\vspace{2em}

Strategic planning is employed to define an organization's direction and the goals that should be pursued in the long run.
Strategic decisions generally concern the whole business instead of individual business units or teams.
In an Agile development context, many strategic decisions are assigned to the Product Owner, which require detailed knowledge of the software product and are primarily concerned with software release plans and product road maps.
However, as the development team possesses deep technical knowledge of the product they should also be involved in defining strategy.

Tactical plans are involved with the management of projects, translating strategic directions into actionable plans for specific organization areas, i.e. how to best structure teams and resources to achieve the given goals.

Operational planning describes the lowest, most concrete level of decisions within an organization.
It defines milestones and success conditions and can detail how, and what portion of, projects and proposals will be tackled in a given operational period.
In an Agile software organization, operational decisions are related to implementing features and ensuring that tasks are carried out efficiently.
An ideal, efficient, Agile development team is involved in shared decision-making not only on the operational but also on the tactical levels~\cite{Moe2012}.

\section{\uppercase{Data in (Agile) Software Development}}
An ever-increasing number of companies, in the domain of software development or otherwise, are relying more on the analyses of collected data to make informed decisions.

\subsection{Business Data}
Businesses have gathered extremely detailed statistics, regarding not only their consumers and users, but also suppliers, alliance partners, and competitors~\cite{Brynjolfsson2011}.
This development is driven by the widespread adoption of enterprise information systems, which collect and store large amounts of business data and enable analyses of these collections.
Examples for these systems are Enterprise Resource Planning (ERP) applications, which present a unified view of the business and contain databases of all business transactions~\cite{umble2003enterprise}.
Recently, analyses and utilization of this collected business data have become more relevant with the introduction of Business Intelligence software solutions, which apply an extensive set of data analytic tools to the operational data of an enterprise~\cite{chen2012business}.
Furthermore, the continuing digitization of society has allowed new possibilities for data collection outside operational business systems~\cite{Brynjolfsson2011}.
Cars, smartphones, home automation systems, and other smart devices that are interacted with on a daily basis are routinely instrumented to capture information regarding their current status and ongoing activities~\cite{Svensson2019}.
This data, collected from ``reality mining'', can be used to recognize social patterns in daily user activity, infer relationships, and, especially relevant for software process improvement, model organizational procedures~\cite{Eagle2006}.
Additionally, clickstream data, user interface interactions and keyword searches from websites and software applications can reveal insights into customer behavior without the need for long-running and costly customer studies~\cite{Brynjolfsson2011}.
These types of detailed as well as diversified data sets are not only generated internally in software-intensive or software development companies but can also be acquired or purchased from third parties, be they public or proprietary~\cite{Provost2013}.
Businesses focused on software development can make use of the available data in these various systems to make informed decisions in their development processes and to stay competitive.
The recent resurgence of machine learning approaches~\cite{Svensson2019}, based on the large amount of available data and promising more automated and powerful data analysis, accelerates these trends.
 
\subsection{Business Data for Software Development}
While large amounts of data are available within software businesses, which could be used to augment decision-making processes, selection and prioritization of product features to be shipped in the next iteration is commonly based on stakeholders' previous experiences, perceptions, and opinions~\cite{Olsson2014,Svensson2019}.
Decisions based on these factors may be inconsistent and, more crucially, lack explanation and links to the evidence and the data that led to them.
These factors make it easier to base decisions on politics or gains for individual teams than to focus on customer value.
Analyses of metrics captured from users of deployed software products, however, enable short feedback cycles between customers and developers.
Instead of guessing and assuming how users will interact with, e.g. a given graphical user interface, developers and stakeholders can observe the actual (mis)usage and base further developments and changes on these insights.
The quality of analytical frameworks and tools being employed is directly related to the quality of decisions derived from them.
However, the characteristics of the descriptions and visualizations that decision makers use, are equally important~\cite{Janssen2017,Matthies2016b}.
In recent research Svensson et al. have pointed out, that while there has been a growing interest in the tools used for data processing, little work has focused on the practitioners' perspective and the context of Agile development~\cite{Svensson2019}.
However, as is the case in the enterprise domain, there is an increased interest in applying machine learning methods and techniques in software engineering contexts to enable higher efficiency~\cite{Feldt2018}.
Supporting the important decisions that are required during software development activities in teams is one of the major application areas for these concepts in the realm of software development processes.

\subsection{Software Project Data}
In addition to data produced by customers, internal business units or third parties, the software development process itself represents a source of valuable data, which is intrinsically highly relevant to making decisions within software companies.
Modern Agile software development relies on the creation, management, and delivery of digital development artifacts~\cite{Fernandez2018}.
Not only is software, i.e. code, produced during regular development activities, but also a range of supporting documents and structures, such as work item descriptions, documentation, and version control information~\cite{Noll2012}.
Software engineers continuously produce data points on their current work and development process~\cite{Ying2005}.
As such, the employed development processes are ``inscribed'' into the produced software artifacts~\cite{de2005seeking}.
The version control system (VCS) which is ubiquitously employed for collaboration in teams, keeps track of the individual changes by developers as well as when the changes were made.
It also captures information on who authored and committed the changes and the goal of the commit is recorded in the commit message~\cite{Santos2016}.
More detailed information on work items might be contained in an issue tracker, such as Jira, with issues detailing the rationale, background, and context of a requested feature~\cite{Ortu2015}.
Frequently employed tools such as Continuous Integration servers or static analysis tools provide data points on the current status and health of the developed software project~\cite{Beller2017,Embury2019effect} .
All of these tools are present in modern Agile teams as necessary prerequisites for efficient communication and collaboration.
However, they also present ``a gold-mine of actionable information''~\cite{Guo2016}, especially in the domain of data-informed decision making in software producing organizations.
There is little overhead in collecting and analyzing this sort of data, as it is already being produced by software teams.
In the case of open.source software development, much of this data is even available publicly~\cite{Linstead2009,Zampetti2017}.
Especially of note for data-informed decision-making is the fact that software project data not only comprehensively documents progress but also provides evidence for failures and problems of the developed product~\cite{Ziftci17}.

\subsection{Decision-Making and Agile Self-Organization}

The concepts of self-organization and self-management underpin the issue of decision-making in Agile teams~\cite{moe2009overcoming}.
Members of a self-organizing Agile team, are ideally responsible not only for working on the tasks they choose to work on, but are also compelled to manage and track their own performance.
Equally, the responsibility for decision-making should be distributed among team members rather than being centralized with few parties~\cite{Moe2012}, see Figure~\ref{fig:decision_levels}.
In Agile teams, the manager's role a the main source of decision-making power is reduced, and developers and product owners---and even customers---may be directly involved in project decisions.
The concept of self-organization directly influences the effectiveness of a team.
The authority for making decisions is delegated to the lowest level of operations, which increases the speed of addressing problems and adapting to changes~\cite{Moe2012}.
While data is available, which can inform decisions regarding development processes, research in this area has also shown that Agile team members rely on their experiences for evaluating design decisions~\cite{Zannier2006}.
Current empirical research offers little clarity on how and why Agile software development teams make business and product-related decisions and whether the team autonomy emphasized in Agile methods leads to teams that can make both strategic and tactical decisions in practice~\cite{Drury2012}.
We hope for, and would like to encourage, more research and exploration of this research field in the future.

\section{\uppercase{Conclusion}}
Agile methods, with their focus on self-organized teams and empirical process control, require customized decision-making processes and procedures in organizations~\cite{Moe2012}.
Agile software development teams, due to the requirement of having to deliver software increments in short iterations, are by definition involved in short-term decision-making~\cite{Svensson2019}.
Managers, in an Agile context, are expected to create an environment that allows team members to make decisions based on the best information available~\cite{Schuh2004}.
However, the question of what attributes define \emph{best information} for a given team is still unanswered and requires additional research.
Furthermore, it is accepted that even if good quality data is available, individuals can ignore analyses or even their own preferences due to rules, traditions or the influence of others~\cite{Dyba2008}.
In this paper, we argue that the principles of data-driven decision making, while still underused, are highly applicable and beneficial to Agile software development.
The combination of data analysis and interpretation by Agile teams of humans can enable better informed decisions, both in the domain of business as well as software development processes.

\balance

\bibliographystyle{apalike}
{\small
\bibliography{library}}

\end{document}